\documentclass[sigconf]{acmart}

\copyrightyear{2021}
\acmYear{2021}
\setcopyright{acmcopyright}
\acmConference[SIGIR '21]{Proceedings of the 44th International ACM SIGIR Conference on Research and Development in Information Retrieval}{July 11--15, 2021}{Virtual Event, Canada}
\acmBooktitle{Proceedings of the 44th International ACM SIGIR Conference on Research and Development in Information Retrieval (SIGIR '21), July 11--15, 2021, Virtual Event, Canada}
\acmPrice{15.00}
\acmISBN{978-1-4503-8037-9/21/07}
\acmDOI{10.1145/3404835.3463117}

\usepackage{bm}
\begin{document}
\fancyhead{}
\title{Deep Position-wise Interaction Network for CTR Prediction}

\author{Jianqiang Huang, Ke Hu, Qingtao Tang, Mingjian Chen, Yi Qi, Jia Cheng, Jun Lei}
\affiliation{%
  \institution{Meituan}
   }
\email{ {huangjianqiang,huke05,tangqingtao,chenmingjian,qiyi02,jia.cheng.sh,leijun}@meituan.com}

\renewcommand{\shortauthors}{Jianqiang and Ke, et al.}

\begin{abstract}
  
Click-through rate(CTR) prediction plays an important role in online advertising and recommender systems. In practice, the training of CTR models depends on click data which is intrinsically biased towards higher positions since higher position has higher CTR by nature.
Existing methods such as actual position training with fixed position inference and inverse propensity weighted training with no position inference alleviate the bias problem to some extend.
However, the different treatment of position information between training and inference will inevitably lead to inconsistency and sub-optimal online performance. Meanwhile, the basic assumption of these methods, i.e., the click probability is the product of  examination probability and relevance probability, is oversimplified and insufficient to model the rich interaction between position and other information. 
In this paper, we propose a \textbf{D}eep \textbf{P}osition-wise \textbf{I}nteraction \textbf{N}etwork (DPIN) to efficiently combine all candidate items and positions for estimating CTR at each position, achieving consistency between offline and online as well as modeling the deep non-linear interaction among position, user, context and item under the limit of serving performance.
Following our new treatment to the position bias in CTR prediction, we propose a new evaluation metrics named PAUC (position-wise AUC) that is suitable for measuring the ranking quality at a given position.
Through extensive experiments on a real world dataset, we show empirically that our method is both effective and efficient in solving position bias problem.
We have also deployed our method in production and observed statistically significant improvement over a highly optimized baseline in a rigorous A/B test. 

\end{abstract}

\ccsdesc[500]{Information systems}
\ccsdesc[500]{Information systems~Sponsored search advertising}
\keywords{CTR Prediction; Position-wise Interaction; Position Bias}


\maketitle

\section{Introduction}

In cost-per-click (CPC) advertising systems, advertisers are charged for every ad click, and advertisements are ranked by the eCPM (effective cost per mile), which is the product of click-through rate (CTR)  and bid price.
Hence,CTR prediction is a core task and has a direct impact in the final revenue and user experience.


In general, the implicit feedback collected from the abundant user clicks is used to train the CTR models.
However, position bias exists in implicit feedback and hurt the model performance.
The position bias happens as users tend to clicks on items in higher position regardless of the items’ actual relevance so that the CTR declines rapidly with the display position\cite{chen2020bias, guo2019pal}. 

Since position signal greatly impacts the CTR prediction, there has been a great deal of work on solving position bias problem. Modeling position as a feature in neural network\cite{ling2017model,zhao2019recommending,haldar2020improving} is widely adopted in industrial applications due to its simplicity and effectiveness, in which actual position feature is added in the wide part of neural network during offline training and a default position value will be used during online inference. To avoid using position feature online, Guo et al.\cite{guo2019pal} proposed a PAL framework to conduct online inference without position information.
Additional, many works use inverse propensity weighting(IPW)\cite{wang2016learning,wang2018position,joachims2017unbiased,ai2018unbiased,yuan2020unbiased,ovaisi2020correcting,agarwal2019estimating,hu2019unbiased} to assign different weights to samples during model training. 
Other methods like knowledge distillation\cite{liu2020general}, adversarial neural networks\cite{moore2018modeling}, pairwise training\cite{beutel2019fairness,jin2020deep} have also been proposed.

Most of the above methods usually assume that the click Bernoulli variable C depends on two hidden Bernoulli variables E and R:
\begin{equation}
p(C=1|u, c, i, k)=p(E=1|k, [s])p(R=1|u, c, i),
\end{equation}
where $p(C=1|u, c, i, k)$ represents the probability that a user $ u $ click on the item $ i $ of $k$-th position in a search context $c$. 
We define the context as a collection of real-time request information including query, search time, location, etc.
$ p(E=1|k, [s]) $ represents the probability that position k is examined, and $[s]$ is the subset of context. Most methods commonly assume that the examination probability depends only on position $k$, i.e., $[s]=\varnothing$. $p(R=1|u, c, i)$ represents the probability that the item is relevant to the user in the context. The assumption is used to eliminate position bias by estimating $ p(E=1|k, [s]) $ implicitly or explicitly(modeling position as a feature in the wide part of neural network can be regarded as an implicit approach). And the relevance probability $ p(R=1|u, c, i) $ is obtained as the predicted CTR. 

The different treatment of position information between training and inference will lead to inconsistency. Furthermore, different users usually have different browsing habits. The examination assumption is oversimplified and the position bias might be related to user, context and item. 

In this paper, we propose a \textbf{D}eep \textbf{P}osition-wise \textbf{I}nteraction \textbf{N}etwork (DPIN) to model $CTR^j_k = p(C=1|u, c, i, k)$  effectively and efficiently, where $ CTR^j_k $ is the predicted CTR of $j$-th candidate item at $k$-th position for a specific request.
The order of items is determined by maximizing $ \sum CTR^j_k bid^j $, which is achieved by a greedy algorithm that selects a most valuable item from top position to bottom position.
The contribution of this paper is summarized as follows:

\begin{itemize}
\item We propose to employ a shallow position-wise combination module with non-linear interaction in DPIN. Given a large number of combinations between candidate items and positions, this module is able to predict all the combinations' CTRs in parallel. This achieves consistency and greatly improves model performance.

\item 
DPIN is the first method, as far as we know, to model user interests in candidate positions. We apply a deep position-wise interaction module to effectively represents deep non-linear interaction among position, user and context.

\item A new evaluation metrics named PAUC (position-wise AUC) is proposed for measuring performance in solving position bias problem. We conduct extensive experiments on a real world dataset to show empirically that DPIN is both effective and efficient.
Online A/B test is also deployed to demonstrate that DPIN has a significant improvement over a highly optimized baseline.

\end{itemize}

\section{Deep Position-wise Interaction Network}

\begin{figure*}[t]
\centering
\vspace{-0.1cm}
\includegraphics[width=0.95\linewidth]{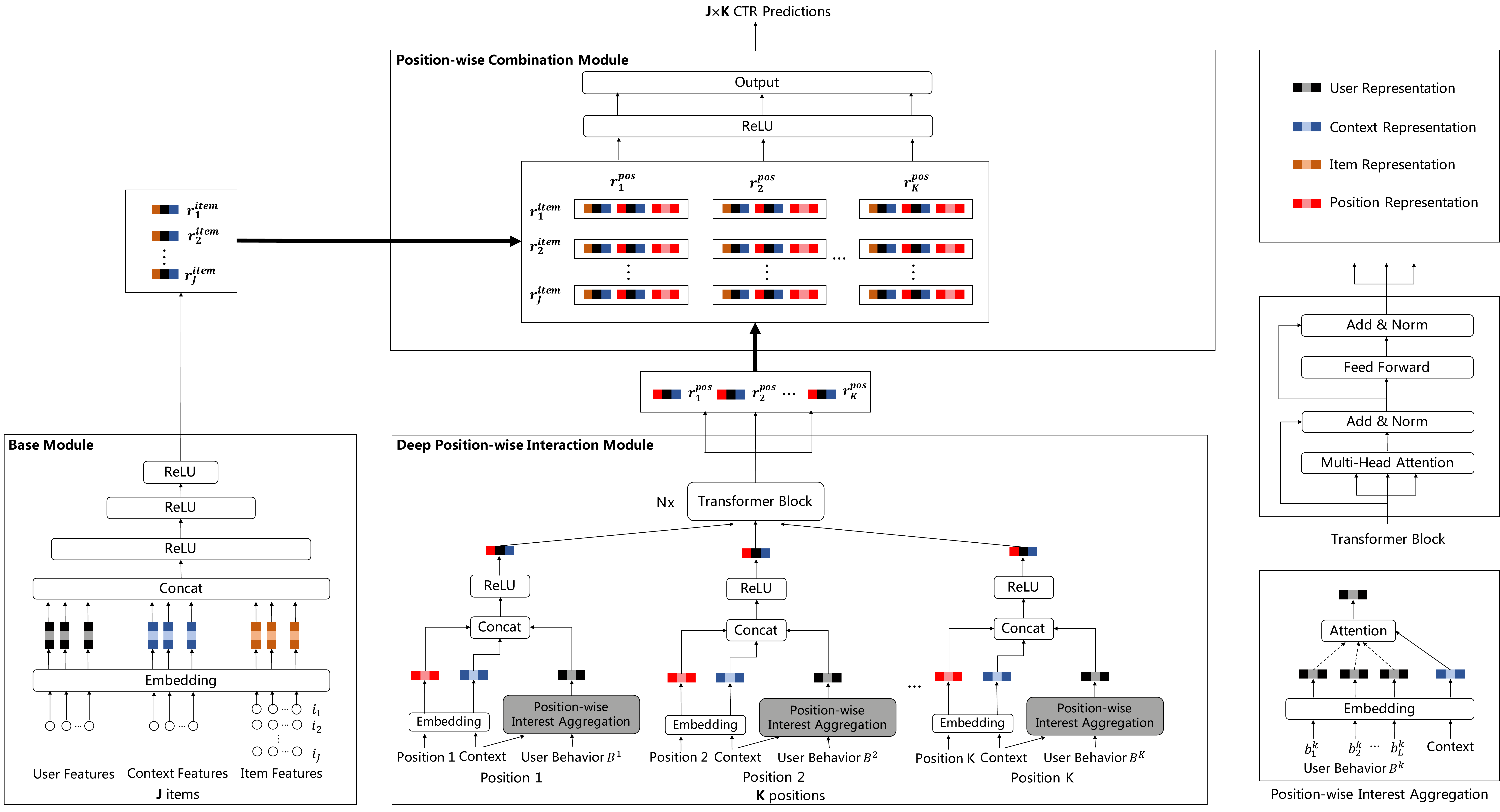}
\caption{The structure of Deep Position-wise Interaction Network.}
\vspace{-0.4cm}
\label{fig:model}

\end{figure*}

In this section, we introduce our DPIN. As shown in Figure \ref{fig:model}, the DPIN is composed of three modules,
which are base module for handling $J$ items, deep position-wise interaction module for handling $K$ positions and position-wise combination module for combining $J$ items and $K$ positions. By this network, CTRs for all candidates at each position can be predicted under the limit of serving performance. We will describe the three modules in detail.


\subsection{Base Module}
Similar to most deep CTR models\cite{cheng2016wide,wang2017deep,guo2017deepfm,zhou2018deep,feng2019deep,zhou2018deep2}, we adopt the structure of embedding and MLP (Multiple Layer Perception) as our base module. For a request, the base module regards an user, a context and $J$ candidate items as inputs, then the representation of j-th item with request information $\bm{r^{item}_{j}}  $ is given via:
\begin{align}
    \label{equ:base-module-u}
    \bm{u} &= Concat(E(u_1),...,E(u_m)),\\
    \label{equ:base-module-c}
    \bm{c} &= Concat(E(c_1),...,E(c_n)),\\
    \label{equ:base-module-ij}
    \bm{i_j} &= Concat(E(i_1^j),...,E(i_o^j)),\\
    \label{equ:base-module-rjitem}
    \bm{r^{item}_{j}} &= MLP(Concat(\bm{u},\bm{c},\bm{i_j})),
\end{align}
where $ \{u_1,...,u_m\}$, $\{c_1,...,c_n\}$,$\{i_1^j,...,i_o^j\}$ are user features set, context features set and j-th item features set respectively. 
$ E(\cdot) \in \mathbb{R}^d $ means the embedding which transforms sparse feature into learnable $d$-dimensional dense vector. Then, the concatenated vector is fed as input of $ MLP $ with $ ReLU (x) = max(0, x) $ activation function.


\subsection{Deep Position-wise Interaction Module}
A brute force method in base module with $O(JKC)$ time complexity for CTR predictions of $K$ positions is unacceptable due to the large time complexity $O(C)$, which is the inference latency of an item at a position.
Therefore, we propose a deep position-wise interaction module paralleled with base module for learning deep non-linear interaction, in which item information is excluded.

In this module, the user behavior sequence at each position is retrieved independently for position-wise interest aggregation to eliminate position bias of lifelong sequence. Then, position-wise non-linear interaction among position, context and user are employed.
Finally, transformer\cite{vaswani2017attention} is adopted for deep interaction among different positions.


\textbf{Position-wise Interest Aggregation.} We represent the user behavior sequence at $k$-th position as $B_k=\{b^k_1,b^k_2,...,b^k_L\}$, in which $ b^k_l=[v^k_l,c^k_l] $ stands for $l$-th behavior record. 
$ v^k_l $ is the $l$-th clicked item(item id, item category are included) and $ c^k_l $ is the context(query, location, hour of day, day of week are included) at which the interaction happened.
Behavior embedding $ \bm{b^k_l} $ can be obtained via:
\begin{equation}
    \bm{b^k_l} = Concat(E(v^{k_l}_1),...,E(v^{k_l}_o),E(c^{k_l}_1),...,E(c^{k_l}_n),E(dif^{k_l})),
\end{equation}
where $ \{v^{k_l}_1,...,v^{k_l}_o\},\{c^{k_l}_1,...,c^{k_l}_n\}  $ are features set of $v^k_l, c^k_l$ respectively and $ dif^{k_l} $ is the time difference between the behavior and current context.

The aggregation of behavior sequence at $k$-th position $ \bm{b_k} $ is calculated by a context-aware attention for extracting user interests related to current context $ \bm{c} $, which can be formulated as follows:
\begin{align}
    \label{equ:behavior aggregation1}
    a_l^k &= ReLu(Concat(\bm{b^k_l}, \bm{c})\bm{W_a} + b_a)\bm{W_b}+b_b, \\
    \label{equ:behavior aggregation2}
    \bm{b_k} &= \sum_{l=1}^L\frac{\exp(a_l^k)}{\sum_{i=1}^{L}\exp (a_i^k)}\bm{b^k_l},
\end{align}
where $ \bm{W_a}, b_a, \bm{W_b}, b_b $ are learnable parameters.

\textbf{Position-wise Non-linear Interaction.} A fully connect with ReLU activation function is employed for non-linear interaction among position, context and user as follows:
\begin{equation}
\bm{v_k} = ReLU(Concat(E(k),\bm{c},\bm{b_k})\bm{W_v}+\bm{b_v}),
\end{equation}
where $ \bm{W_v}, \bm{b_v} $ map concatenated vector to $ d_{model} $ dimension.

\textbf{Transformer Block.} If $ \bm{v_k} $ is regarded as a non-linear interaction representation of $k$-th position, user interests at other positions will be lost. Hence the transformer is adopted for interaction among different positions. Same as \cite{vaswani2017attention}, we donate the input of transformer as $ \bm{Q}=\bm{K}=\bm{V}= Concat(\bm{v_1},\bm{v_2},...,\bm{v_K}) \in \mathbb{R}^{K \times d_{model}} $.  Multi-head self-attention can be formulated as follows:
\begin{align}
    \label{equ:multihead1}
    MultiHead(\bm{Q}, \bm{K}, \bm{V}) &= Concat(\bm{head_1},...,\bm{head_h})\bm{W^O}, \\
    \label{equ:multihead2}
   \bm{head_i} &= Attention(\bm{QW^Q_i},\bm{KW^K_i},\bm{VW^V_i}),\\
    \label{equ:multihead3}
    Attention(\bm{Q,K,V}) &= softmax(\frac{\bm{QK^T}}{\sqrt{d_k}})\bm{V},
\end{align}
where $ \bm{W^Q_i}, \bm{W^K_i}, \bm{W^V_i} \in \mathbb{R}^{d_{model} \times d_k},\bm{W^O} \in \mathbb{R}^{d_{model} \times d_{model}}$ are the parameter matrices. And $ d_k = d_{model}/h $ is the dimension of each head. The output of multi-head self-attention is fed as the input of position-wise feed-forward network for non-linear transformation.
Residual connections and layer normalization are adopted successively. And a stack of $ N $ transformer blocks is employed for deep interaction.

Finally, the representation of $k$-th position with request information $ \bm{r_k^{pos}} $ is given by the proposed deep position-wise interaction module.




\subsection{Position-wise Combination Module}
The purpose of the position-wise combination module is to predict the CTR of each item at each position by combining J items and K positions. A non-linear interaction among $ \bm{r_j^{item}} $, $ \bm{r_k^{pos}} $ and $ E(k) $ is used to learn non-linear relationship among user, context, item and position. The CTR of j-th item at k-th position $ CTR_k^j $ can be calculated as follows:
\begin{small}
\begin{equation}
CTR^j_k = \sigma(ReLU(Concat(\bm{r_j^{item}},\bm{r_k^{pos}}, E(k))\bm{W_1} + \bm{b_1})\bm{W_2} + b_2),
\end{equation}
\end{small}
where $ \bm{W_1}, \bm{b_1} $ are the parameters of non-linear interaction layer. $ \bm{W_2} , b_2  $ are the parameters of output layer. $ \sigma(\cdot) $ is sigmoid function.

Finally, the model can be trained using stochastic gradient descent algorithm with actual position feature, and cross-entropy is used as the loss function.

\section{Experiment}

In this section, we evaluate the model performance and serving performance of the proposed DPIN. We describe the experimental settings and experimental results in detail.
\subsection{Experimental Settings}
\textbf{Datasets.} A four-week ad impression log collected from a sponsored search advertising system in a shopping application is used to train the CTR models. We evaluate our methods on two test sets, which are collected from regular traffic and top-k randomized traffic the next day. The top-k randomized traffic is suitable for position bias evaluation since it excludes the impact of relevance recommendations. The number of impressions is about 10 million a day, and 5\% of traffic is the randomized traffic. 

\textbf{Metrics.} We use AUC (Area Under ROC) as one of our evaluation metrics. To evaluate the model performance of our methods for position bias, we propose PAUC (Position-wise AUC) as another evaluation metrics, which is calculated as follows:
\begin{equation}
PAUC=\frac{
\sum_{k=1}^{K}\#impression_kPAUC@k}{\sum_{k=1}^{K}\#impression_k},
\end{equation}
where $\#impression_k$ is the number of impressions at $k$-th position, and $PAUC@k$ is the AUC of impressions at $k$-th position. PAUC measures the relevance ranking quality at every position, ignoring the impact of position bias.

\textbf{Compared Methods.} Our baseline is a highly optimized DIN\cite{zhou2018deep} model, in which a large number of attributes and hand-crafted features are added. The number of features is up to 241 and embedding dimension $k$ is 8 for each feature. The hidden sizes of $MLP$ are 1024, 512, 128. And the hidden size of the non-linear interaction layer in position-wise combination module is 128. The length of users' behaviors is truncated to 300. And the number of positions $K$ is truncated to 25, which accounts for most of the online traffic. In order to compare different methods fairly, we ensure that all input information and parameters setting of common module are consistent. 
We conducted experiments with the following methods:

\textbf{DIN.} Position is not used in this model.

\textbf{DIN+PosInWide.} The method models position feature in the wide part of neural network and first position is used for evaluation.

\textbf{DIN+PAL.} PAL method is adopted based on DIN.

\textbf{DIN+ActualPosInWide.} The method use actual position for evaluation based on DIN+PosInWide.

\textbf{DIN+Combination.} Position-wise combination module is employed based on DIN. Actual position is used for evaluation.

\textbf{DPIN-Transformer.} Transformer is not used in DPIN.

\textbf{DPIN.} There is the proposed DPIN model, in which a tow-layer transformer is adopted and $d_{model}=64,h=2$ for self-attention.

\textbf{DPIN+ItemAction.} 
We add deep position-wise interaction module before the MLP layer of the base module, and candidate item information is introduced in position-wise interest aggregation and position-wise non-linear interaction. The experiment is the upper bound of model performance in our methods, but the serving performance is unacceptable.

\subsection{Offline Evaluations}

\begin{table}[ht]
\caption{Offline experimental results of compared methods on the regular and randomized test sets.}
\label{table:OfflineEvaluations}
\begin{center}
\begin{small}
\begin{sc}
\begin{tabular}{lcccccr}
\toprule
&\multicolumn{2}{c}{$Regular$} & \multicolumn{2}{c}{$Randomized$} \\
Model & AUC &  PAUC & AUC &  PAUC\\
\midrule
DIN & 0.7818 &  0.7090 & 0.7836 & 0.7223\\
DIN+PosInWide & 0.7696 &  0.7109 & 0.7725 & 0.7239 \\
DIN+PAL & 0.7735 &  0.7128 & 0.7763 & 0.7254 \\
DIN+AcutalPosInWide & 0.7928 & 0.7109 &  0.7938 &  0.7239 \\
DIN+Combination   & 0.7970 & 0.7172 & 0.7985 & 0.7294 \\
DPIN-Transformer & 0.7961 & 0.7148 &  0.7984 & 0.7283 \\
\textbf{DPIN} & \textbf{0.7994} & \textbf{0.7216} &  \textbf{0.8015} & \textbf{0.7350} \\
DPIN+ItemAction & 0.7999 & 0.7223 & 0.8019 &  0.7356\\
\bottomrule
\end{tabular}
\end{sc}
\end{small}
\end{center}
\vskip -0.1in
\end{table}

Table \ref{table:OfflineEvaluations} illustrates experimental results of the compared methods on both regular and randomized test sets. We first analyze the differences between different methods on the regular traffic. 
Compared with DIN, the DIN+PosInWide and DIN+PAL have a performance degradation on AUC but an improvement on PAUC, which shows that both methods effectively alleviate position bias but lead to inconsistency between offline and online.
The DIN+AcutalPosInWide solves the inconsistency by introducing actual position during evaluation, which is achievable by position-wise combination module. But modeling position in the wide part leads to the position feature is only a bias, which can not improve PAUC. 
By employing our proposed position-wise combination module, the DIN+Combination has 2.74\% gain on AUC and 0.63\% gain on PAUC compared with DIN+PosInWide, achieving consistency and alleviate position bias further, which shows the position bias is not independent.
Furthermore, DPIN models deep non-linear interaction among position, context and user, and eliminates position bias existed in the user sequence by position-wise method, which has 0.24\% gain on AUC and 0.44\% gain on PAUC compared with DIN+Combination. 
The effect of DPIN-Transformer explains that it is necessary to adopt transformer for interaction among different positions.
And the comparison between DPIN and DPIN+ItemAction shows that DPIN is close to the brute force method on both AUC and PAUC.
As can be seen finally, the DPIN has 2.98\% gain on AUC and 1.07\% gain on PAUC relative to the DIN+PosInWide, which is a baseline in the advertising system online. 

In order to ensure that our method can learn the position bias instead of overfitting the selection bias of the system, we further evaluate our methods on randomized traffic. 
The results show that the differences between the different methods on both regular and randomized traffic is consistent.




\subsection{Serving Performance}
\begin{figure}
\vspace{-0.5cm}
\centering
\setlength{\abovecaptionskip}{0.2cm}

\includegraphics[width=0.95\linewidth]{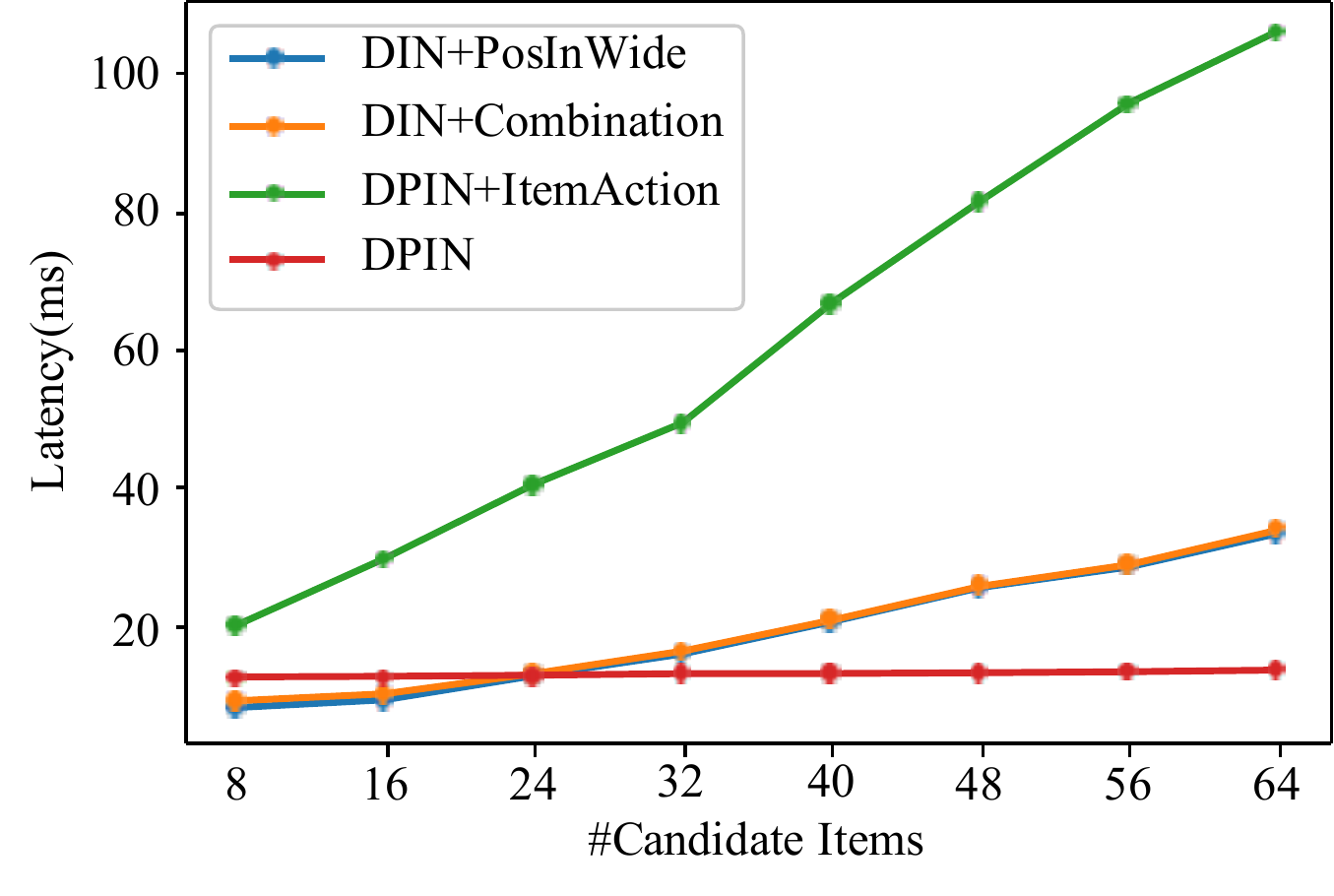}
\caption{Serving latency changes with the number of candidate items in different methods.}

\vspace{-0.6cm}
\label{fig:serving}
\end{figure}
We retrieve some requests with different candidate item numbers from the dataset to measure serving performance. As shown in Figure \ref{fig:serving}, the serving latency of position-wise combination module is negligible compared to the DIN model since user sequence operation has a large proportion of latency.
The serving latency of DPIN increases slowly as the number of items increases since the deep position-wise interaction module has nothing to do with items. Compared with DIPIN+ItemAction, 
the DPIN has a great improvement in serving performance with little damage to model performance, which shows that our proposed method is both effective and efficient.

\subsection{Online Evaluations}
Online A/B test was conducted in the sponsored search advertising system from 2021-01-08 to 2021-01-22. For the control group, 10\% of users are randomly selected and presented with recommendation generated by DIN+PosInWide. For the experimental group, 10\% of users are presented with recommendation generated by DPIN. The A/B test shows that the proposed DPIN has improved CTR by 2.25\% and RPM (Revenue Per Mille) by 2.15\% compared with baseline. For now, DPIN has been deployed online and serves the main traffic, which contributes a significant business revenue growth.

\section{CONCLUSIONS}
In this paper, a novel model Deep Position-wise Interaction Network (DPIN) is proposed to mitigate position bias, which efficiently combine all candidate items and positions for estimating CTR at each position, achieving consistency between offline and online. At the same time, the deep non-linear interaction among position, context and user is available by the model. In order to evaluate our method in position bias problem, we propose a new evaluation metrics PAUC and the offline experiments show that the proposed DPIN outperforms the compared methods efficacy and efficiency. For now, the DPIN is deployed in a sponsored search advertising system and serving the main traffic.

\bibliographystyle{ACM-Reference-Format}


\begin{thebibliography}{24}


\ifx \showCODEN    \undefined \def \showCODEN     #1{\unskip}     \fi
\ifx \showDOI      \undefined \def \showDOI       #1{#1}\fi
\ifx \showISBNx    \undefined \def \showISBNx     #1{\unskip}     \fi
\ifx \showISBNxiii \undefined \def \showISBNxiii  #1{\unskip}     \fi
\ifx \showISSN     \undefined \def \showISSN      #1{\unskip}     \fi
\ifx \showLCCN     \undefined \def \showLCCN      #1{\unskip}     \fi
\ifx \shownote     \undefined \def \shownote      #1{#1}          \fi
\ifx \showarticletitle \undefined \def \showarticletitle #1{#1}   \fi
\ifx \showURL      \undefined \def \showURL       {\relax}        \fi
\providecommand\bibfield[2]{#2}
\providecommand\bibinfo[2]{#2}
\providecommand\natexlab[1]{#1}
\providecommand\showeprint[2][]{arXiv:#2}

\bibitem[\protect\citeauthoryear{Agarwal, Zaitsev, Wang, Li, Najork, and
  Joachims}{Agarwal et~al\mbox{.}}{2019}]%
        {agarwal2019estimating}
\bibfield{author}{\bibinfo{person}{Aman Agarwal}, \bibinfo{person}{Ivan
  Zaitsev}, \bibinfo{person}{Xuanhui Wang}, \bibinfo{person}{Cheng Li},
  \bibinfo{person}{Marc Najork}, {and} \bibinfo{person}{Thorsten Joachims}.}
  \bibinfo{year}{2019}\natexlab{}.
\newblock \showarticletitle{Estimating position bias without intrusive
  interventions}. In \bibinfo{booktitle}{\emph{Proceedings of the Twelfth ACM
  International Conference on Web Search and Data Mining}}.
  \bibinfo{pages}{474--482}.
\newblock


\bibitem[\protect\citeauthoryear{Ai, Bi, Luo, Guo, and Croft}{Ai
  et~al\mbox{.}}{2018}]%
        {ai2018unbiased}
\bibfield{author}{\bibinfo{person}{Qingyao Ai}, \bibinfo{person}{Keping Bi},
  \bibinfo{person}{Cheng Luo}, \bibinfo{person}{Jiafeng Guo}, {and}
  \bibinfo{person}{W~Bruce Croft}.} \bibinfo{year}{2018}\natexlab{}.
\newblock \showarticletitle{Unbiased learning to rank with unbiased propensity
  estimation}. In \bibinfo{booktitle}{\emph{The 41st International ACM SIGIR
  Conference on Research \& Development in Information Retrieval}}.
  \bibinfo{pages}{385--394}.
\newblock


\bibitem[\protect\citeauthoryear{Beutel, Chen, Doshi, Qian, Wei, Wu, Heldt,
  Zhao, Hong, Chi, et~al\mbox{.}}{Beutel et~al\mbox{.}}{2019}]%
        {beutel2019fairness}
\bibfield{author}{\bibinfo{person}{Alex Beutel}, \bibinfo{person}{Jilin Chen},
  \bibinfo{person}{Tulsee Doshi}, \bibinfo{person}{Hai Qian},
  \bibinfo{person}{Li Wei}, \bibinfo{person}{Yi Wu}, \bibinfo{person}{Lukasz
  Heldt}, \bibinfo{person}{Zhe Zhao}, \bibinfo{person}{Lichan Hong},
  \bibinfo{person}{Ed~H Chi}, {et~al\mbox{.}}} \bibinfo{year}{2019}\natexlab{}.
\newblock \showarticletitle{Fairness in recommendation ranking through pairwise
  comparisons}. In \bibinfo{booktitle}{\emph{Proceedings of the 25th ACM SIGKDD
  International Conference on Knowledge Discovery \& Data Mining}}.
  \bibinfo{pages}{2212--2220}.
\newblock


\bibitem[\protect\citeauthoryear{Chen, Dong, Wang, Feng, Wang, and He}{Chen
  et~al\mbox{.}}{2020}]%
        {chen2020bias}
\bibfield{author}{\bibinfo{person}{Jiawei Chen}, \bibinfo{person}{Hande Dong},
  \bibinfo{person}{Xiang Wang}, \bibinfo{person}{Fuli Feng},
  \bibinfo{person}{Meng Wang}, {and} \bibinfo{person}{Xiangnan He}.}
  \bibinfo{year}{2020}\natexlab{}.
\newblock \showarticletitle{Bias and Debias in Recommender System: A Survey and
  Future Directions}.
\newblock \bibinfo{journal}{\emph{arXiv preprint arXiv:2010.03240}}
  (\bibinfo{year}{2020}).
\newblock


\bibitem[\protect\citeauthoryear{Cheng, Koc, Harmsen, Shaked, Chandra, Aradhye,
  Anderson, Corrado, Chai, Ispir, et~al\mbox{.}}{Cheng et~al\mbox{.}}{2016}]%
        {cheng2016wide}
\bibfield{author}{\bibinfo{person}{Heng-Tze Cheng}, \bibinfo{person}{Levent
  Koc}, \bibinfo{person}{Jeremiah Harmsen}, \bibinfo{person}{Tal Shaked},
  \bibinfo{person}{Tushar Chandra}, \bibinfo{person}{Hrishi Aradhye},
  \bibinfo{person}{Glen Anderson}, \bibinfo{person}{Greg Corrado},
  \bibinfo{person}{Wei Chai}, \bibinfo{person}{Mustafa Ispir}, {et~al\mbox{.}}}
  \bibinfo{year}{2016}\natexlab{}.
\newblock \showarticletitle{Wide \& deep learning for recommender systems}. In
  \bibinfo{booktitle}{\emph{Proceedings of the 1st workshop on deep learning
  for recommender systems}}. \bibinfo{pages}{7--10}.
\newblock


\bibitem[\protect\citeauthoryear{Feng, Lv, Shen, Wang, Sun, Zhu, and Yang}{Feng
  et~al\mbox{.}}{2019}]%
        {feng2019deep}
\bibfield{author}{\bibinfo{person}{Yufei Feng}, \bibinfo{person}{Fuyu Lv},
  \bibinfo{person}{Weichen Shen}, \bibinfo{person}{Menghan Wang},
  \bibinfo{person}{Fei Sun}, \bibinfo{person}{Yu Zhu}, {and}
  \bibinfo{person}{Keping Yang}.} \bibinfo{year}{2019}\natexlab{}.
\newblock \showarticletitle{Deep session interest network for click-through
  rate prediction}.
\newblock \bibinfo{journal}{\emph{arXiv preprint arXiv:1905.06482}}
  (\bibinfo{year}{2019}).
\newblock


\bibitem[\protect\citeauthoryear{Guo, Tang, Ye, Li, and He}{Guo
  et~al\mbox{.}}{2017}]%
        {guo2017deepfm}
\bibfield{author}{\bibinfo{person}{Huifeng Guo}, \bibinfo{person}{Ruiming
  Tang}, \bibinfo{person}{Yunming Ye}, \bibinfo{person}{Zhenguo Li}, {and}
  \bibinfo{person}{Xiuqiang He}.} \bibinfo{year}{2017}\natexlab{}.
\newblock \showarticletitle{DeepFM: a factorization-machine based neural
  network for CTR prediction}.
\newblock \bibinfo{journal}{\emph{arXiv preprint arXiv:1703.04247}}
  (\bibinfo{year}{2017}).
\newblock


\bibitem[\protect\citeauthoryear{Guo, Yu, Liu, Tang, and Zhang}{Guo
  et~al\mbox{.}}{2019}]%
        {guo2019pal}
\bibfield{author}{\bibinfo{person}{Huifeng Guo}, \bibinfo{person}{Jinkai Yu},
  \bibinfo{person}{Qing Liu}, \bibinfo{person}{Ruiming Tang}, {and}
  \bibinfo{person}{Yuzhou Zhang}.} \bibinfo{year}{2019}\natexlab{}.
\newblock \showarticletitle{PAL: a position-bias aware learning framework for
  CTR prediction in live recommender systems}. In
  \bibinfo{booktitle}{\emph{Proceedings of the 13th ACM Conference on
  Recommender Systems}}. \bibinfo{pages}{452--456}.
\newblock


\bibitem[\protect\citeauthoryear{Haldar, Ramanathan, Sax, Abdool, Zhang,
  Mansawala, Yang, Turnbull, and Liao}{Haldar et~al\mbox{.}}{2020}]%
        {haldar2020improving}
\bibfield{author}{\bibinfo{person}{Malay Haldar}, \bibinfo{person}{Prashant
  Ramanathan}, \bibinfo{person}{Tyler Sax}, \bibinfo{person}{Mustafa Abdool},
  \bibinfo{person}{Lanbo Zhang}, \bibinfo{person}{Aamir Mansawala},
  \bibinfo{person}{Shulin Yang}, \bibinfo{person}{Bradley Turnbull}, {and}
  \bibinfo{person}{Junshuo Liao}.} \bibinfo{year}{2020}\natexlab{}.
\newblock \showarticletitle{Improving Deep Learning For Airbnb Search}. In
  \bibinfo{booktitle}{\emph{Proceedings of the 26th ACM SIGKDD International
  Conference on Knowledge Discovery \& Data Mining}}.
  \bibinfo{pages}{2822--2830}.
\newblock


\bibitem[\protect\citeauthoryear{Hu, Wang, Peng, and Li}{Hu
  et~al\mbox{.}}{2019}]%
        {hu2019unbiased}
\bibfield{author}{\bibinfo{person}{Ziniu Hu}, \bibinfo{person}{Yang Wang},
  \bibinfo{person}{Qu Peng}, {and} \bibinfo{person}{Hang Li}.}
  \bibinfo{year}{2019}\natexlab{}.
\newblock \showarticletitle{Unbiased lambdamart: an unbiased pairwise
  learning-to-rank algorithm}. In \bibinfo{booktitle}{\emph{The World Wide Web
  Conference}}. \bibinfo{pages}{2830--2836}.
\newblock


\bibitem[\protect\citeauthoryear{Jin, Fang, Zhang, Ren, Zhou, Xu, Yu, Wang,
  Zhu, and Gai}{Jin et~al\mbox{.}}{2020}]%
        {jin2020deep}
\bibfield{author}{\bibinfo{person}{Jiarui Jin}, \bibinfo{person}{Yuchen Fang},
  \bibinfo{person}{Weinan Zhang}, \bibinfo{person}{Kan Ren},
  \bibinfo{person}{Guorui Zhou}, \bibinfo{person}{Jian Xu},
  \bibinfo{person}{Yong Yu}, \bibinfo{person}{Jun Wang},
  \bibinfo{person}{Xiaoqiang Zhu}, {and} \bibinfo{person}{Kun Gai}.}
  \bibinfo{year}{2020}\natexlab{}.
\newblock \showarticletitle{A deep recurrent survival model for unbiased
  ranking}. In \bibinfo{booktitle}{\emph{Proceedings of the 43rd International
  ACM SIGIR Conference on Research and Development in Information Retrieval}}.
  \bibinfo{pages}{29--38}.
\newblock


\bibitem[\protect\citeauthoryear{Joachims, Swaminathan, and Schnabel}{Joachims
  et~al\mbox{.}}{2017}]%
        {joachims2017unbiased}
\bibfield{author}{\bibinfo{person}{Thorsten Joachims}, \bibinfo{person}{Adith
  Swaminathan}, {and} \bibinfo{person}{Tobias Schnabel}.}
  \bibinfo{year}{2017}\natexlab{}.
\newblock \showarticletitle{Unbiased learning-to-rank with biased feedback}. In
  \bibinfo{booktitle}{\emph{Proceedings of the Tenth ACM International
  Conference on Web Search and Data Mining}}. \bibinfo{pages}{781--789}.
\newblock


\bibitem[\protect\citeauthoryear{Ling, Deng, Gu, Zhou, Li, and Sun}{Ling
  et~al\mbox{.}}{2017}]%
        {ling2017model}
\bibfield{author}{\bibinfo{person}{Xiaoliang Ling}, \bibinfo{person}{Weiwei
  Deng}, \bibinfo{person}{Chen Gu}, \bibinfo{person}{Hucheng Zhou},
  \bibinfo{person}{Cui Li}, {and} \bibinfo{person}{Feng Sun}.}
  \bibinfo{year}{2017}\natexlab{}.
\newblock \showarticletitle{Model ensemble for click prediction in bing search
  ads}. In \bibinfo{booktitle}{\emph{Proceedings of the 26th International
  Conference on World Wide Web Companion}}. \bibinfo{pages}{689--698}.
\newblock


\bibitem[\protect\citeauthoryear{Liu, Cheng, Dong, He, Pan, and Ming}{Liu
  et~al\mbox{.}}{2020}]%
        {liu2020general}
\bibfield{author}{\bibinfo{person}{Dugang Liu}, \bibinfo{person}{Pengxiang
  Cheng}, \bibinfo{person}{Zhenhua Dong}, \bibinfo{person}{Xiuqiang He},
  \bibinfo{person}{Weike Pan}, {and} \bibinfo{person}{Zhong Ming}.}
  \bibinfo{year}{2020}\natexlab{}.
\newblock \showarticletitle{A general knowledge distillation framework for
  counterfactual recommendation via uniform data}. In
  \bibinfo{booktitle}{\emph{Proceedings of the 43rd International ACM SIGIR
  Conference on Research and Development in Information Retrieval}}.
  \bibinfo{pages}{831--840}.
\newblock


\bibitem[\protect\citeauthoryear{Moore, Pfeiffer, Wei, Iyer, Charles,
  Gilad-Bachrach, Boyles, and Manavoglu}{Moore et~al\mbox{.}}{2018}]%
        {moore2018modeling}
\bibfield{author}{\bibinfo{person}{John Moore}, \bibinfo{person}{Joel
  Pfeiffer}, \bibinfo{person}{Kai Wei}, \bibinfo{person}{Rishabh Iyer},
  \bibinfo{person}{Denis Charles}, \bibinfo{person}{Ran Gilad-Bachrach},
  \bibinfo{person}{Levi Boyles}, {and} \bibinfo{person}{Eren Manavoglu}.}
  \bibinfo{year}{2018}\natexlab{}.
\newblock \showarticletitle{Modeling and Simultaneously Removing Bias via
  Adversarial Neural Networks}.
\newblock \bibinfo{journal}{\emph{arXiv preprint arXiv:1804.06909}}
  (\bibinfo{year}{2018}).
\newblock


\bibitem[\protect\citeauthoryear{Ovaisi, Ahsan, Zhang, Vasilaky, and
  Zheleva}{Ovaisi et~al\mbox{.}}{2020}]%
        {ovaisi2020correcting}
\bibfield{author}{\bibinfo{person}{Zohreh Ovaisi}, \bibinfo{person}{Ragib
  Ahsan}, \bibinfo{person}{Yifan Zhang}, \bibinfo{person}{Kathryn Vasilaky},
  {and} \bibinfo{person}{Elena Zheleva}.} \bibinfo{year}{2020}\natexlab{}.
\newblock \showarticletitle{Correcting for selection bias in learning-to-rank
  systems}. In \bibinfo{booktitle}{\emph{Proceedings of The Web Conference
  2020}}. \bibinfo{pages}{1863--1873}.
\newblock


\bibitem[\protect\citeauthoryear{Vaswani, Shazeer, Parmar, Uszkoreit, Jones,
  Gomez, Kaiser, and Polosukhin}{Vaswani et~al\mbox{.}}{2017}]%
        {vaswani2017attention}
\bibfield{author}{\bibinfo{person}{Ashish Vaswani}, \bibinfo{person}{Noam
  Shazeer}, \bibinfo{person}{Niki Parmar}, \bibinfo{person}{Jakob Uszkoreit},
  \bibinfo{person}{Llion Jones}, \bibinfo{person}{Aidan~N Gomez},
  \bibinfo{person}{Lukasz Kaiser}, {and} \bibinfo{person}{Illia Polosukhin}.}
  \bibinfo{year}{2017}\natexlab{}.
\newblock \showarticletitle{Attention is all you need}.
\newblock \bibinfo{journal}{\emph{arXiv preprint arXiv:1706.03762}}
  (\bibinfo{year}{2017}).
\newblock


\bibitem[\protect\citeauthoryear{Wang, Fu, Fu, and Wang}{Wang
  et~al\mbox{.}}{2017}]%
        {wang2017deep}
\bibfield{author}{\bibinfo{person}{Ruoxi Wang}, \bibinfo{person}{Bin Fu},
  \bibinfo{person}{Gang Fu}, {and} \bibinfo{person}{Mingliang Wang}.}
  \bibinfo{year}{2017}\natexlab{}.
\newblock \showarticletitle{Deep \& cross network for ad click predictions}.
\newblock In \bibinfo{booktitle}{\emph{Proceedings of the ADKDD'17}}.
  \bibinfo{pages}{1--7}.
\newblock


\bibitem[\protect\citeauthoryear{Wang, Bendersky, Metzler, and Najork}{Wang
  et~al\mbox{.}}{2016}]%
        {wang2016learning}
\bibfield{author}{\bibinfo{person}{Xuanhui Wang}, \bibinfo{person}{Michael
  Bendersky}, \bibinfo{person}{Donald Metzler}, {and} \bibinfo{person}{Marc
  Najork}.} \bibinfo{year}{2016}\natexlab{}.
\newblock \showarticletitle{Learning to rank with selection bias in personal
  search}. In \bibinfo{booktitle}{\emph{Proceedings of the 39th International
  ACM SIGIR conference on Research and Development in Information Retrieval}}.
  \bibinfo{pages}{115--124}.
\newblock


\bibitem[\protect\citeauthoryear{Wang, Golbandi, Bendersky, Metzler, and
  Najork}{Wang et~al\mbox{.}}{2018}]%
        {wang2018position}
\bibfield{author}{\bibinfo{person}{Xuanhui Wang}, \bibinfo{person}{Nadav
  Golbandi}, \bibinfo{person}{Michael Bendersky}, \bibinfo{person}{Donald
  Metzler}, {and} \bibinfo{person}{Marc Najork}.}
  \bibinfo{year}{2018}\natexlab{}.
\newblock \showarticletitle{Position bias estimation for unbiased learning to
  rank in personal search}. In \bibinfo{booktitle}{\emph{Proceedings of the
  Eleventh ACM International Conference on Web Search and Data Mining}}.
  \bibinfo{pages}{610--618}.
\newblock


\bibitem[\protect\citeauthoryear{Yuan, Liu, Hsia, Dong, and Lin}{Yuan
  et~al\mbox{.}}{2020}]%
        {yuan2020unbiased}
\bibfield{author}{\bibinfo{person}{Bowen Yuan}, \bibinfo{person}{Yaxu Liu},
  \bibinfo{person}{Jui-Yang Hsia}, \bibinfo{person}{Zhenhua Dong}, {and}
  \bibinfo{person}{Chih-Jen Lin}.} \bibinfo{year}{2020}\natexlab{}.
\newblock \showarticletitle{Unbiased Ad click prediction for position-aware
  advertising systems}. In \bibinfo{booktitle}{\emph{Fourteenth ACM Conference
  on Recommender Systems}}. \bibinfo{pages}{368--377}.
\newblock


\bibitem[\protect\citeauthoryear{Zhao, Hong, Wei, Chen, Nath, Andrews,
  Kumthekar, Sathiamoorthy, Yi, and Chi}{Zhao et~al\mbox{.}}{2019}]%
        {zhao2019recommending}
\bibfield{author}{\bibinfo{person}{Zhe Zhao}, \bibinfo{person}{Lichan Hong},
  \bibinfo{person}{Li Wei}, \bibinfo{person}{Jilin Chen},
  \bibinfo{person}{Aniruddh Nath}, \bibinfo{person}{Shawn Andrews},
  \bibinfo{person}{Aditee Kumthekar}, \bibinfo{person}{Maheswaran
  Sathiamoorthy}, \bibinfo{person}{Xinyang Yi}, {and} \bibinfo{person}{Ed
  Chi}.} \bibinfo{year}{2019}\natexlab{}.
\newblock \showarticletitle{Recommending what video to watch next: a multitask
  ranking system}. In \bibinfo{booktitle}{\emph{Proceedings of the 13th ACM
  Conference on Recommender Systems}}. \bibinfo{pages}{43--51}.
\newblock


\bibitem[\protect\citeauthoryear{Zhou, Mou, Fan, Pi, Bian, Zhou, Zhu, and
  Gai}{Zhou et~al\mbox{.}}{2018a}]%
        {zhou2018deep2}
\bibfield{author}{\bibinfo{person}{Guorui Zhou}, \bibinfo{person}{Na Mou},
  \bibinfo{person}{Ying Fan}, \bibinfo{person}{Qi Pi}, \bibinfo{person}{Weijie
  Bian}, \bibinfo{person}{Chang Zhou}, \bibinfo{person}{Xiaoqiang Zhu}, {and}
  \bibinfo{person}{Kun Gai}.} \bibinfo{year}{2018}\natexlab{a}.
\newblock \showarticletitle{Deep Interest Evolution Network for Click-Through
  Rate Prediction}.
\newblock \bibinfo{journal}{\emph{arXiv preprint arXiv:1809.03672}}
  (\bibinfo{year}{2018}).
\newblock


\bibitem[\protect\citeauthoryear{Zhou, Zhu, Song, Fan, Zhu, Ma, Yan, Jin, Li,
  and Gai}{Zhou et~al\mbox{.}}{2018b}]%
        {zhou2018deep}
\bibfield{author}{\bibinfo{person}{Guorui Zhou}, \bibinfo{person}{Xiaoqiang
  Zhu}, \bibinfo{person}{Chenru Song}, \bibinfo{person}{Ying Fan},
  \bibinfo{person}{Han Zhu}, \bibinfo{person}{Xiao Ma},
  \bibinfo{person}{Yanghui Yan}, \bibinfo{person}{Junqi Jin},
  \bibinfo{person}{Han Li}, {and} \bibinfo{person}{Kun Gai}.}
  \bibinfo{year}{2018}\natexlab{b}.
\newblock \showarticletitle{Deep interest network for click-through rate
  prediction}. In \bibinfo{booktitle}{\emph{Proceedings of the 24th ACM SIGKDD
  International Conference on Knowledge Discovery \& Data Mining}}.
  \bibinfo{pages}{1059--1068}.
\newblock


\end{thebibliography}
\balance 

\appendix

\end{document}